\documentstyle[11pt,fleqn]{article}  
\catcode`\@=11
\@addtoreset{equation}{section}
\def\theequation{\arabic{section}.\arabic{equation}}
\def\appendix{\renewcommand{\thesection}{\Alph{section}}\setcounter{section}{0}
              \renewcommand{\theequation}
            {\mbox{\Alph{section}.\arabic{equation}}}\setcounter{equation}{0}}
\def\maketitle{\thispagestyle{empty}\setcounter{page}0\newpage
                \renewcommand{\thefootnote}{\arabic{footnote}}
                  \setcounter{footnote}0}
\renewcommand{\thanks}[1]{\renewcommand{\thefootnote}{\fnsymbol{footnote}}
               \footnote{#1}\renewcommand{\thefootnote}{\arabic{footnote}}}
\newcommand{\preprint}[1]{\hfill{\sl preprint - #1}\par\bigskip\par\rm}
\renewcommand{\title}[1]{\begin{center}\Large\bf #1\end{center}\rm\par\bigskip}
\renewcommand{\author}[1]{\begin{center}\Large #1\end{center}}
\newcommand{\address}[1]{\begin{center}\large #1\end{center}}

\newcommand{\pacs}[1]{\smallskip\noindent{\sl PACS number(s):
                       \hspace{0.3cm}#1}\par\bigskip\rm}
\def\babs{\hrule\par\begin{description}\item{Abstract: }\it} 
\def\eabs{\par\end{description}\hrule\par\medskip\rm}
\renewcommand{\date}[1]{\par\bigskip\par\sl\hfill #1\par\medskip\par\rm}
\newcommand{\ack}[1]{\par\section*{Acknowledgments} #1} 
\newcommand{\s}[1]{\section{#1}}

\renewcommand{\vec}[1]{{\bf #1}}       
\def\hs{\qquad\qquad}         
\def\nn{\nonumber}            
\def\beq{\begin{eqnarray}}    
\def\eeq{\end{eqnarray}}      
\def\ap{\left.}               
\def\at{\left(}               
\def\aq{\left[}               
\def\ag{\left\{}              
\def\cp{\right.}              
\def\ct{\right)}              
\def\cq{\right]}              
\def\cg{\right\}}             
\def\R{{\hbox{{\rm I}\kern-.2em\hbox{\rm R}}}}   
\def\H{{\hbox{{\rm I}\kern-.2em\hbox{\rm H}}}}   
\def\N{{\hbox{{\rm I}\kern-.2em\hbox{\rm N}}}}   
\def\C{{\ \hbox{{\rm I}\kern-.6em\hbox{\bf C}}}} 
\def\Z{{\hbox{{\rm Z}\kern-.4em\hbox{\rm Z}}}}   
\def\ii{\infty}                                  
\newcommand{\fr}[2]{\mbox{$\frac{#1}{#2}$}}      
\def\Det{\,\mbox{Det}\,}                
\def\Tr{\,\mbox{Tr}\,}                  
\def\al{\alpha}
\def\be{\beta}

\def\ze{\zeta}

\def\si{\sigma}
\def\om{\omega}

\def\Ga{\Gamma}

\def\Om{\Omega}

\def\citen#1{%
\edef\@tempa{\@ignspaftercomma,#1, \@end, }
\edef\@tempa{\expandafter\@ignendcommas\@tempa\@end}%
\if@filesw \immediate \write \@auxout {\string \citation {\@tempa}}\fi
\@tempcntb\m@ne \let\@h@ld\relax \let\@citea\@empty
\@for \@citeb:=\@tempa\do {\@cmpresscites}%
\@h@ld}
\def\@ignspaftercomma#1, {\ifx\@end#1\@empty\else
   #1,\expandafter\@ignspaftercomma\fi}
\def\@ignendcommas,#1,\@end{#1}
\def\@cmpresscites{%
 \expandafter\let \expandafter\@B@citeB \csname b@\@citeb \endcsname
 \ifx\@B@citeB\relax 
    \@h@ld\@citea\@tempcntb\m@ne{\bf ?}%
    \@warning {Citation `\@citeb ' on page \thepage \space undefined}%
 \else
    \@tempcnta\@tempcntb \advance\@tempcnta\@ne
    \setbox\z@\hbox\bgroup 
    \ifnum\z@<0\@B@citeB \relax
       \egroup \@tempcntb\@B@citeB \relax
       \else \egroup \@tempcntb\m@ne \fi
    \ifnum\@tempcnta=\@tempcntb 
       \ifx\@h@ld\relax 
          \edef \@h@ld{\@citea\@B@citeB}%
       \else 
          \edef\@h@ld{\hbox{--}\penalty\@highpenalty \@B@citeB}%
       \fi
    \else   
       \@h@ld \@citea \@B@citeB \let\@h@ld\relax
 \fi\fi%
 \let\@citea\@citepunct
}
\def\@citepunct{,\penalty\@highpenalty\hskip.13em plus.1em minus.1em}%
\def\@citex[#1]#2{\@cite{\citen{#2}}{#1}}%
\def\@cite#1#2{\leavevmode\unskip
  \ifnum\lastpenalty=\z@ \penalty\@highpenalty \fi 
  \ [{\multiply\@highpenalty 3 #1
      \if@tempswa,\penalty\@highpenalty\ #2\fi 
    }]\spacefactor\@m}
\begin{document}
\input{epsf}

\preprint{DFTUZ/95/29 and UTF-366}

\title
{Interaction of Low - Energy Induced Gravity 
\\ with Quantized Matter -- II. Temperature effects}

\author{G. Cognola\thanks{e-mail address: cognola@science.unitn.it}}
\address{Dipartimento di Fisica, Universit\`a di Trento and
\\ Istituto Nazionale di Fisica Nucleare, Gruppo Collegato di Trento, Italia}

\author{I. L. Shapiro\thanks{e-mail address: shapiro@dftuz.unizar.es}}
\address{Departamento de Fisica Teorica\\
Universidad de Zaragoza, 50009, Zaragoza, Spain and
\\ Tomsk State Pedagogical Institute, Tomsk, 634041, Russia }

\date{December 1995}

\babs
At the very early Universe the matter fields are described
by the GUT models in curved space-time. At high energies these fields
are asymptotically free and conformally coupled to external metric. The only
possible quantum effect is the appearance of the conformal anomaly, which
leads to the propagation of the new degree of freedom - conformal factor.
Simultaneously with the expansion of the Universe, the scale of energies
decreases and the propagating conformal factor starts to interact with 
the Higgs field due to the violation of conformal invariance in the matter 
fields sector. In a previous paper \cite{foo} we have shown that this
interaction can lead to special physical effects like the renormalization 
group flow, which ends in some fixed point. 
Furthermore in the vicinity of this fixed point there occur the first order 
phase transitions. In the present paper we consider the same theory
of conformal factor coupled to Higgs field and incorporate the temperature
effects. We reduce the complicated higher-derivative operator 
to several ones of the standard second-derivative form and calculate 
an exact effective potential with temperature on the 
anti de Sitter (AdS) background. 
The physical analysis of the effective potential is performed 
in the framework of the high-temperature expansion.
\eabs

\maketitle

\s{Introduction}
The increasing interest to the quantum field theory in an external 
classical gravitational field is based on the following reasons.
First of all such an approach is usually regarded as a first step 
towards the more complete theory of quantum gravity. On the other 
hand, while the issue of quantum gravity itself is related to the physics 
at Planck scale of energies, the effects of an external gravitational 
background can be relevant at the energies lower than the Planck ones and 
therefore it can be described in more familiar theoretical models
\cite{davies,fulling,book}. In particular, renormalization and 
the notion of renormalization group can be introduced for arbitrary
curved background and this enables one to study asymptotical behaviour,
phase transitions, vacuum structure and other aspects of quantum matter 
fields. For instance, if we are dealing with the GUT scale, then all the
interactions are supposed to be asymptotically free. Furthermore,
for some GUT models there is a specific phenomena of asymptotic
conformal invariance \cite{buch83-25,book} and as a consequence at high 
energies we meet the quantum matter fields free of all the interactions
except the conformal one with an external metric. The requirement 
of conformal invariance mainly concerns the scalar (Higgs) field, 
which fixes the value of the nonminimal parameter of scalar - curvature 
interaction to the conformal value 1/6. 

 From the cosmological point of view, the stage when the matter fields
are described by the Grand Unification
Theories corresponds to inflation \cite{guth81-23,lind82-108}. 
During the inflationary epoch the Universe expands exponentially 
in time simultaneously with the
decreasing of the character energy scale of all the interactions. 
Then at the lower energies the
conformal invariance in the Higgs sector gets violated and as a result
the scalar fields start to interact with the conformal factor of the
metric. 
In a previous paper \cite{foo} (see also \cite{induce}) we have explored 
some physical
consequences of this interaction. In particular it was shown that the 
theory of scalars interacting with the conformal factor
is renormalizable and that the renormalization group flow ends in the
infra-red (IR) stable fixed point with the minimally coupled scalar field
free of four-scalar interactions. Furthermore, the effects of quantum 
conformal factor lead to a new kind of first order phase transition 
induced by curvature, where the Higgs field plays the role of order parameter
(see also \cite{eliz93-315} where the potential of conformal
factor itself was explored).
Thus we have estimated the quantum effects of conformal
factor to the physics of matter fields, that is in fact a new specific 
way of taking into account the back reaction of vacuum to the 
matter felds. 

In all the considerations, we have used the standard perturbative rules
which are equivalent to the zero - temperature approximation. At the same
time for the sake of completeness one must take into account the temperature 
effects as well, since they could have dominated in the early Universe
\cite{kirz72-42}. 
In the present paper we report about the study of the temperature effects
in the theory of induced gravity coupled to the matter fields. Since the
divergences of the theory are not affected by the temperature,
we can use the renormalization group flows established in Ref.~\cite{foo} 
for the zero-temperature case and consider the physical
effects in the vicinity of the fixed points. However, the temperature 
effects lead to a strong modifications in the effective potential and
so the physical results are essentially different in the present case.

The paper is organized as follows. In section 2 we remind the 
basical statements of the previous paper \cite{foo} and in particular we 
write down the renormalization group equations for the effective couplings.
Section 3 is devoted to the calculation of the effective potential in our
theory of dilaton coupled to matter in the high-temperature regime on 
a space-time with  constant curvature. We choose the AdS background
because it enables one to formulate the equilibrium state and consider the
temperature effects in a consistent way 
\cite{hawk83-87-577,alle87-189-304} (see also Ref~\cite{vanz94-50-5148}
and references cited therein). 
 We overcome the difficulties related with the higher derivatives which 
occur in the action of conformal factor by the use of some simple 
transformations which are especially useful for the calculation of the 
effective potential.
As a result we obtain the general exact expression for the effective 
potential which depends on a lot of arbitrary parameters including temperature,
curvature, point of normalization for the Higgs field and scaling constant.
Thus the general expression is not suitable for analysis and we are enforced
to make some simplifications. In order to do this in section 4 we consider 
the special high temperature case and find that in this limit, the most of 
the above mentioned arbitrariness disappear.

\s{Induced gravity and its interaction with matter} 
\label{S:IG}

Let us start with some asymptotic free and asymptotically conformal 
invariant GUT in curved space-time. 
The multiplicative renormalizability of the GUT model in curved space-time 
requires the nonminimal terms of the form $\xi R\phi^2$
to be included into the action of every Higgs field $\phi$,
as well as the vacuum terms. 
When the radiational corrections are taken into account, the parameter
of nonminimal coupling obeys the corresponding renormalization - group 
equations\footnote{We consider the case of one real scalar field 
for simplicity and therefore we deal with the single nonminimal parameter.
In general, the situation may be more complicated.}.
As it was discovered in \cite{buch83-25} (see also \cite{book}),
some asymptotic free models are also asymptotically
conformal invariant.
This means that the nonminimal coupling $\xi$
is arbitrary at low energies while at high energies it tends to the special
conformal value ${1}/{6}$. Below we consider only this class of
gauge models. Since all the masses also vanish in far ultra-violet 
(UV) regime, at that high
scale we are effectively living with free massless fields conformally 
coupled to the metric background.
Since the interaction between 
matter fields are weakened then the only quantum effect is the appearance 
of the trace anomaly of the energy-momentum tensor.
The purpose of the present paper is to explore the back reaction
of this vacuum effect to the matter fields, taking into account 
temperature effects, which could have been important in the Early Universe. 

The anomaly trace
of the energy-momentum tensor appears due to the divergences and the lack
of completely invariant regularization \cite{dese76-111} (see \cite{20let}
for the complete references). It the theory under
discussion the anomalous trace is the combination of the vacuum beta-functions.
Since the anomaly is known it enables one to calculate, with accuracy to
some conformal invariant functional, the effective action of vacuum 
\cite{reig84-134,frad84-134}
(see also \cite{book} and \cite{foo}).
The effective action originally arises as a nonlocal functional,
but it can be written in a local form with the help
of an extra dimensionless field $\sigma$ which can be named as dilaton,
in analogy with the string theory or 
as conformal factor, according to its origin. It reads
\beq
W[g_{\mu\nu},\sigma] = S_{c}[g_{\mu\nu}]+
\int d^4x\sqrt{-g}\left\{\fr12\sigma\Delta\sigma
+\sigma\left[k'_1C^2+k'_2\left(E-\fr23\Box R\right)\right]
+k'_3R^2\right\}
\label{EAc}\:.\eeq
Here the conformally covariant self-adjoint operator $\Delta$
is defined by
\beq
\Delta=\Box^{2}+
2R^{\mu\nu}\nabla_{\mu}\nabla_{\nu}-\fr23R\Box
+\fr13(\nabla^{\mu}R)\nabla_{\mu}
\label{4}\:.\eeq
The values of $k'_{1,2,3}$ are defined by the amount of the fields of
spin $0$, $\frac12$ and  $1$ in starting GUT theory 
and for our purposes they are not relevant, as well as the conformal 
invariant nonlocal action $S_{c}[g_{\mu\nu}]$.

Our main supposition \cite{foo} is that the quantum effects of induced gravity,
that is of the field $\sigma$,
are relevant below the scale of 
asymptotic conformal invariance, where the nonminimal parameter $\xi$
runs away from the conformal value. We are interested in the cosmological
applications and therefore it is natural to suppose that the transition to
low energies (or long distances) corresponds to some conformal transformation
in the induced gravity action (\ref{EAc}) and hence the classical fields and
induced gravity appear in different conformal 
points\footnote{Very recently the supersymmetric 
generalization of Ref. \cite{anto92-45} has been performed   
in Ref. \cite{buch9511205}.} \cite{anto92-45,induce,foo}. 
Thus it is necessary to make a conformal transformation of the metric 
in (\ref{EAc}) and then consider the unified theory. 
At the same time it is much more convenient to make the
conformal transformation of metric and matter fields in the action of the last.
Such a transformation corresponds to some change of variables in the path
integral for the unified theory.

The only source of conformal noninvariance in the action of 0,
$\frac{1}{2}$ and 1 spin fields
is the nonminimal term in the scalar sector.
In the framework of asymptotically conformal invariant models the value of
$\xi$ is not equal to $\frac{1}{6}$ at low energies
and hence the interaction of conformal factor with scalar field arises.
Introducing the scale parameter $\alpha$ we obtain the following action
for the conformal factor coupled with the scalar field, that is
\beq
S&=&W[g_{\mu\nu},\sigma]+\int d^{4}x \sqrt{-g}
\left\{
\fr12(1-6\xi)\phi^2\left(\alpha^2
g^{\mu\nu}\partial_{\mu}\sigma\partial_{\nu}\sigma
+\alpha\Box\sigma\right)
\right.\nn\\&&\hs\hs\hs\left.
+\fr12g^{\mu\nu}\partial_{\mu}\phi\partial_{\nu}\phi
+\fr12\xi R\phi^2-\fr1{24}f\phi^4\right\}
\label{5}\:,\eeq
where $W[g_{\mu\nu},\sigma]$ is defined iby Eq.~(\ref{EAc}).
Thus the interaction between scalar field and conformal factor
arises as a result of the conformal transformation of the metric
$g_{\mu\nu}\to g'_{\mu\nu}=g_{\mu\nu}\exp(2\alpha\sigma)$
and the matter field
$\Phi\rightarrow \Phi'=\Phi\exp(d_{\Phi}\alpha\sigma)$,
where $d_{\Phi}$ is the conformal weight of the field
$\Phi$. The only kind of fields which takes part in such an interaction is
the scalar one,
where the interaction with conformal factor appears as a result of
nonconformal coupling at low energies.
Hence the contributions of other matter fields to the
effective potential are not important, since they can
only change the values of $k'_{1,2,3}$ in Eq.~(\ref{EAc}).

We must take into account the quantum effects of conformal factor coupled to 
ordinary scalar fields and to estimate the quantum corrections to the 
effective potential of the last. For the sake of quantum calculations 
we shall follow \cite{foo} and use the background field method.
We separate the fields into background
$\sigma,\phi$ and quantum $\tau,\eta$ ones, by means
\beq
\sigma\rightarrow \sigma' = \sigma + \tau \:,\hs
\phi \rightarrow \phi' = \phi + \eta
\label{6}\:.\eeq

The one-loop contribution to the effective action 
is defined as (see Sec.~\ref{S:OLEP})
\beq
\Gamma^{(1)}=\frac{1}{2}\Tr\ln{\hat{H}}=\frac12\ln\Det\hat H
\label{7}\:,\eeq
where ${\hat{H}}$ is the bilinear
(with respect to quantum fields $\tau,\eta$) form of the classical action 
(\ref{5}):
\beq
\hat{H}=\left(\matrix{
H_{\tau\tau} &H_{\tau\eta}\cr
H_{\eta\tau} &H_{\eta\eta}\cr} \right)
\label{8}\:,\eeq
where
\beq
H_{\tau\tau} = \Box^{2}+
2R^{\mu\nu}\nabla_{\mu}\nabla_{\nu}-\fr23R\Box
+\fr13(\nabla^{\mu}R)\nabla_{\mu}
+(6\xi-1)\left[\alpha^2\phi^2\Box
+\alpha^2(\nabla^{\mu}\phi^2)\nabla_{\mu}\right]
\:,\nn\eeq
\beq
H_{\tau\eta} =
-(6\xi-1)\left[\al\phi\Box+2\al(\nabla^\mu\phi)\nabla_\mu
+\al(\Box\phi)
-2\al^2\phi(\nabla^\mu\si)\nabla_\mu
-2\al^2(\nabla_\mu(\phi\nabla^\mu\si))\right]
\:,\nn\eeq
\beq
H_{\eta\tau}=-(6\xi-1)\left[\al\phi\Box
+2\al^2\phi(\nabla^\mu\si)\nabla_\mu\right]
\:,\nn\eeq
\beq
H_{\eta\eta} =-\Box+\xi R-\fr12f\phi^2
-(6\xi-1)\left[\alpha(\Box\sigma)
+\alpha^2(\nabla^\mu\sigma)
(\nabla_\mu\sigma)\right]\:.
\label{9}\eeq
The one-loop divergences show that
the theory (\ref{5}) is (at least at one-loop) 
renormalizable\footnote{To construct the completely 
renormalizable theory one probably needs to
use a more general metric-dilaton action introduced in \cite{eli}.}
and therefore one can use the renormalization group method for its study.
At this stage it is more convenient to deal with
an arbitrary background metric $g_{\mu\nu}$ and thus
we use the approach described in \cite{book}.
The general solution of the renormalization group
equations for effective action
\beq
\left\{ \mu\frac{d}{d\mu}+\beta_f \frac{d}{d f}
+\beta_\xi \frac{d}{d \xi}
-\gamma_{\phi}\frac{\delta}{\delta \phi}
- \gamma_{\sigma}\frac{\delta}{\delta \sigma}
\right\}\Gamma[\phi,\sigma,f,\xi,g_{\mu\nu},\mu]=0
\label{15}\eeq
has the form
\beq
\Gamma[\phi, \sigma,f,\xi,g_{\mu\nu}e^{2t}, \mu] =
\Gamma [\phi(t),\sigma(t),f(t),\xi(t),g_{\mu\nu},\mu]
\label{16}\:,\eeq
where $\mu$ is the dimensional parameter of renormalization.
Effective coupling constants obey the equations \cite{foo}
\beq
(4\pi)^2\frac{df(t)}{dt'} = \beta_f = 3f^2 + 12f\alpha^2 \zeta^2 + 
12\alpha^4 \zeta^2{(\zeta - 1)}^2,
\hs f(0) = f\:,
\nn\eeq
\beq
(4\pi)^2\frac{d\zeta(t)}{dt'} = \beta_\zeta = \zeta [f +
 2\alpha^2 \zeta (\zeta - 1)],
\hs \xi(0) = \xi \:,
\label{17}\eeq
while $\gamma$-functions are equal to zero. 
Here we use $\zeta(t) = 1-6\xi(t)$ instead of $\xi$ for compactness.

The study of asymptotics of the effective couplings
$f(t), \zeta(t)$ can be done with the fixed point method. 
It is easy to see that both $\beta$ functions (\ref{17}) vanish
in the physically relevant points $f = 0, \zeta = 0$ and $f = 0, \zeta = 1$.
There are also three more solutions with negative $f$, but we do not discuss 
them here. The first solution corresponds to the conformal fixed point while
the second one corresponds to the minimal fixed point with $\xi = 0$.
The analysis of stability of
the minimal fixed point has been done in the framework of standard 
Lyapunov's method and it was found to be stable in the IR limit 
$t \rightarrow - \infty$. Moreover in this limit
$f\sim\xi^6\rightarrow 0$ and therefore $f\ll\xi$. We remark that such a
"far" IR regime for the gravitational effects 
under discussion is in fact an UV regime for 
the matter fields interactions and hence we can consider only the 
effects related with the  
conformal factor as relevant. It should not be so if $f$ is the
same order as $\zeta$. The conformal fixed points turns 
out to be unstable \cite{foo} and it plays the role of initial point for 
the renormalization group flow. 

The renormalization group equations (\ref{16}) and  (\ref{17})
enable one to restore the effective action of scalar field $\phi$ with
any preassigned accuracy \cite{book}. However since we need to estimate 
the temperature corrections, which are not caused by
divergences, it is much more useful to use the $\zeta$-regularization technique
choosing the special AdS background. 
As we shall see in the next section, in this way we shall reduce 
the case of higher derivative operator (\ref{8}) to the
case of second order operators for which known results are at 
disposal.

\s{Finite temperature 
one-loop effective potential 
in AdS space.} \label{S:OLEP}

Here we briefly recall how is defined the finite temperature 
one-loop effective potential within the path integral approach to 
quantum field theory. 
In a static gravitational background, the finite temperature theory
is obtained by the compactification, in the imaginary time 
$\tau=ix^0$, of the Euclidean section of the manifold. In this way 
$0<\tau<\be$ ($\be=1/T$ being the inverse temperature) and the fields
have to satisfy periodic boundary conditions, that is 
$\phi(\tau,\vec x)=\phi(\tau+\be,\vec x)$.  

Let us consider the simple case of ordinary scalar field $\phi$ on the 
background of an external metric.
A functional Taylor expansion of the action around the classical solution
$\hat\phi$ gives 
\begin{equation}
S[\phi,g]=S_c[\hat\phi,g]
+\left.\frac{\delta^2S[\phi,g]}{\delta\phi^2}\right|_{\phi=\hat\phi}
\frac{\eta^2}{2} +\mbox{ higher order terms in }\eta,
\label{BGexp}\end{equation} 
$S_c[\hat\phi,g]$ being the classical action and
so the one-loop approximated theory is determined by the partition function 
\begin{equation}
Z[\hat\phi,g]\sim\exp(-S_c[\hat\phi,g])\int d[\eta]\,
\exp\left(-\frac{1}{2}\int\eta A\eta\,d^4x\right).
\label{PI0}\end{equation}
Since we have performed a Wick rotation of the time axis in
the complex plane, the metric $g_{\mu\nu}$ is Euclidean and the 
small disturbance operator $A$ selfadjoint and non-negative.
With these assumptions, the partition function can be formally
computed in terms of the real eigenvalues of the operator
$A$ \cite{hawk77-55-133} and for the one-loop quantum corrections 
to the classical action one obtains 
\beq
\Ga^{(1)}[\hat\phi,g]=\frac12\ln\Det\frac{A}{\mu^2}
\:,\label{logDet}\eeq
where $\mu$ is an arbitrary mass, which is necessary to adjust 
dimensions. It will be determined by renormalization. 
The latter equation is valid also for a multiplet of matter fields.
In such a case $A$ is a matrix of differential operators and one gets
Eq.~(\ref{7}). 

The effective action can be written in the 
form (see for example Ref.~\cite{aber73-9-1})
\beq
\Gamma[\hat\phi,g]=\int\left[V(\hat\phi,g)+\frac{1}{2}Z(\hat\phi,g)
g^{ij}\partial_i\hat\phi\partial_j\hat\phi+\cdots\right]\sqrt{g}\,d^4x
\:.\label{OLEA}\eeq 
The latter equation implicitly defines also
the contributions to the one-loop effective potential $V^{(1)}(\hat\phi,g)$.
Using $\zeta$-function for the definition of the determinant in 
Eq.~(\ref{logDet}), we finally get
\beq
V^{(1)}(\hat\phi,g)=-\ap\frac12\ze'\at0;x|A/\mu^2\ct
\right|_{\hat\phi=const}
\:,\eeq
where $\ze(s;x|A)$ is the $\zeta$-function density related to the 
operator $A$.

In our case, the small disturbance operator ($\hat H$) is quite 
complicated, but it notably simplifies in a maximally symmetric background. 
For this reason we choose the anti de Sitter space-time,
where the equilibrium state at any temperature is well defined.
Now, we briefly resume some known results in AdS, refering the reader
to the literature for more details (see for example
\cite{fron65-37-221,hawk83-87-577,alle87-189-304,camp91-43-3958,
camp92-45-3591,vanz94-50-5148}).

Let us start by considering a scalar field in AdS 
satisfying the Klein-Gordon equation
\beq
\at-\Box+M^2-\frac94\Om^2\ct\phi=0
\:,\label{11n}\eeq
where $M\geq3\Om/2$ is a constant and $\Om^2$ is proportional to the 
scalar curvature. We have
\beq
R_{\mu\nu}=-3\Om^2g_{\mu\nu}\:,\hs
R=-12\Om^2
\:.\label{12n}\eeq

For this case, the zero temperature contribution 
$V^{(1)}_0$ to one-loop effective 
potential can be computed by observing that the Euclidean section 
of AdS is the hyperbolic manifold $H^4$ and on such a 
space the $\zeta$-function is explicitly known 
\cite{camp91-43-3958,byts94u-325}.
So one has
\beq
V^{(1)}_0(M)&=&-\frac1{64\pi^2}\aq
M^4\at\frac32-\ln\frac{M^2}{\mu^2}\ct
-\frac{M^2\Om^2}2\at1-\ln\frac{M^2}{\mu^2}\ct
\cq\nn\\
&&\hs\hs-\frac{\Om^4}{8\pi^2}\int_0^\ii
\frac{x(x^2+1/4)}{e^{2\pi x}+1}
\:\ln\at\frac{M^2+\Om^2x^2}{\mu^2}\ct\:dx
\label{V0}
\:.\eeq 
The arbitrary mass scale $\mu$ has to be fixed by renormalization 
\cite{burg85-153-137,cogn94-50-909}. 
The small curvature expansion of the latter expression reads
\beq
V^{(1)}_0(M)&\sim&
-\frac1{64\pi^2}\aq
M^4\at\frac32-\ln\frac{M^2}{\mu^2}\ct
-\frac{M^2}2\at1-\ln\frac{M^2}{\mu^2}\ct\:\Om^2
\cp\nn\\&&\ap\hs\hs
+\frac{17}{240}\ln\frac{M^2}{\mu^2}\:\Om^4
\cq+O(\Om^4)
\label{V0app}\:.\eeq

The finite temperature contribution $V^{(1)}_T$ has also been 
already computed directly using the thermodynamical formulae. 
It can be written in the form \cite{vanz94-50-5148}
\beq
V^{(1)}_T(M)=-\frac{\Om^4}{i\pi^3}\int_{c-i\ii}^{c+i\ii}
\Ga(1-s)\ze_R(s)\chi(s-1;\om_0)\at\frac\Om{T}\ct^{-s}\, ds
\:,\label{VT}\eeq
where $c>4$, $\om_0=M/\Om+3/2$ and 
\beq
\chi(s;\om_0)&=&\frac14\aq\ze_H(s-2,\om_0)
-2\at\om_0-\frac32\ct\ze_H(s-1,\om_0)
\cq\nn\\&&\ap\hs\hs
+(\om_0-1)(\om_0-2)\ze_H(s,\om_0)\cq
\:,\eeq
$\ze_R(s)$ and $\ze_H(s,\om_0)$ being the Riemann and Hurwitz 
$\zeta$-functions respectively.

This representation of the effective potential is very useful
in the computation of high temperature expansion, which we need. 
In fact, we can compute the integral by the residues method, by
observing that the integral function has simple poles at the points
$s=4,3,2,0,-1-2,...$ and a double pole at $s=1$. 
In this way we obtain an asymptotic expansion for the potential,
since the contribution of the contour integral at the (left) infinity,
which we disregard, is exponentially vanishing in $T$.
One easily obtains \cite{vanz94-50-5148}
\beq
V^{(1)}_T(M)&\sim&-\ag
\frac{\pi^2}{90}T^4
-\frac{\ze_R(3)M}\pi T^3
+\frac1{12}\at M^2-\frac{\Om^2}4\ct T^2 
\cp\nn\\&&\ap
+\frac2\pi\aq
\chi(0;\om_0)\ln\frac{T}\Om+\chi'(0;\om_0)
\cq\Om^3T
+\frac{\chi(-1;\om_0)}{\pi}\Om^4
\cg +O\at\frac1T\ct
\:.\label{VTapp}\eeq
\vskip 3mm

In our case the action assumes the form (\ref{5})
and the one-loop effective action is defined according 
to (\ref{7}), with the bilinear form (\ref{9}).
The  expressions (\ref{9}) can be notably simplified,  
because we are interested in the effective potential
for the scalar $\phi$ 
(this means that  $\phi$ and $\si$ are constant) 
in AdS
(that is $R=-12\Om^2=const$ and $R_{\mu\nu}=Rg_{\mu\nu}/4$).
Then, the operator we are dealing with assumes the form 
\begin{eqnarray}
\hat{H}=\left(\matrix{\Box^2+ U\Box  
& x \Box \cr
x \Box 
&-\Box + D \cr} \right)
\label{Hoper}\:,\end{eqnarray}
where
\beq
U = (6\xi-1)\al\phi + 2\Om^2,\hs
x = (1-6\xi)\al\phi,\hs  
D = - 12\xi \Om^2 -\frac f2\phi^2
 \label{num}\eeq
can be considered as some numbers since nor $\phi$ neither $\Omega$
do not depend on the spacetime variables. Consequently, any element
of the matrix (\ref{Hoper}) commutes with each other and
the finite temperature one-loop effective action can be obtained 
without special calculations taking into account the standard results 
written in the previous section.

To see this we observe that
\beq
2\Gamma^{(1)} &=& \ln\Det\hat{H} = 
\ln\Det\left(\matrix{\Box & 0 \cr 
                        0 & -1\cr}\right)
+\ln\Det\left(\matrix{\Box+U & x     \cr
                      -x\Box & \Box-D\cr} \right) 
\nn\\&=&
\ln\Det(-\Box)+\ln\Det\at\Box-D\ct+\ln\Det\left(
\Box+U+x^2\Box(\Box-D)^{-1}\Box\right)
\label{14n}\:.\end{eqnarray}
Since all the operators commute the last expression can be reduced to
\begin{eqnarray}
2\Gamma = 
\ln\Det(-\Box)+\ln\Det(z_1-\Box)+
\ln\Det(z_2-\Box)
\label{15n}\:,\end{eqnarray}
where 
\beq
z_{1,2} = - \frac12 (U - D + x^2) 
\pm \sqrt{\frac14(U - D + x^2)^2 + UD}
\eeq 
are the roots of the equation
\beq
z^2 + (U - D + x^2)z - UD = 0
\:.\eeq
Now, using the above results for the 
effective potential $V_{eff}$  we immediately get
\beq
V_{eff} = V_{cl}+V^{(1)}_0 + V^{(1)}_T
\eeq
where 
\beq
V_{cl}=-\xi R\phi^2/2+f\phi^4/24
\eeq 
is the classical potential and
\beq
V^{(1)}_0= 
V^{(1)}_0(M_0)+V^{(1)}_0(M_1)+V^{(1)}_0(M_2)
\:,\eeq
\beq
V^{(1)}_T = V^{(1)}_T(M_0)+V^{(1)}_T(M_1)+V^{(1)}_T(M_2)
\:,\label{VT123}
\eeq
the expressions for $V^{(1)}_0(M)$ and $V^{(1)}_T(M)$ being given by 
Eqs.~(\ref{V0}) and (\ref{VT}) respectively, with
\beq
M_0^2=9\Om^2/4,\hs M_1^2=z_1+9\Om^2/4,\hs 
M_2^2=z_2+9\Om^2/4
\:.\label{Ms}\eeq
The above expression gives an exact one-loop effective potential with 
temperature on the AdS background for our model of induced gravity coupled to 
matter fields. Below we discuss some physical applications of this result.

\s{First order phase transitions of the system}

First order phase transitions of the system can happen for some critical 
values of the parameters, say $\phi_c,T_c,R_c$, which are 
determined by the equations
\beq
&&V_{eff}(\phi_c,T_c,R_c)-V_{eff}(0,T_c,R_c)=0\:,\nn\\
&&V'_{eff}(\phi_c,T_c,R_c)=0\:,\hs
V''_{eff}(\phi_c,T_c,R_c)>0
\:.\label{phase}\eeq
Higher loop contributions are significant in the determination of
the exact critical parameters \cite{dola74-9-3320} then, for consistency, 
in the determination of the critical values, we have 
to use the approximated expressions, Eqs.~(\ref{V0app}) and 
(\ref{VTapp}), in the computation of the effective potential.

Let us now consider the form of the effective potential in the vicinity of
the IR stable fixed point $f = 0$, $\xi = 0$. As it was already 
mentioned, the 
effective couplings tend to these values in a different way. In particular,
$f \sim \xi^6 \rightarrow 0$. 
In practice, in the vicinity of
the fixed point we shall keep only the lower powers of the small parameter
$\xi$ and therefore omit $f$ almost everywhere. Indeed one must be careful
in expanding the nonlinear functions like the square roots in the series of 
small parameter like $\xi$, since for big values of the scalar field, the 
first terms of the expansion can behave different as compared with original 
function. Next, we shall suppose that the system is in the high temperature 
phase and that $T>>\Omega$.
Using the RG based restrictions $f\sim\xi^6\rightarrow0$ 
into the zero-temperature effective potential, one obtains the 
renormalized expression \cite{foo}, which satisfies the same normalization 
conditions (we keep them in what follows).
\beq
V_{cl}+V_0&=&6\xi\Omega^2\phi^2 
+\frac{3}{16\pi^2}\;(1 - 6\xi)^2\alpha^2\Omega^2\phi^2 
\left[\ln\frac{\phi^2}{\mu^2}-3\right]
\nn\\&&\hs\hs
+\frac{9}{16\pi^2} \;\xi^2 (1 - 6\xi)^2 \alpha^4 
 \phi^4\left[\ln\frac{\phi^2}{\mu^2}
-\frac{25}6\right]                                   
\:.\label{1n}\eeq
With this restrictions, the quantities in Eq.~(\ref{Ms}) simplify to
\beq
M_0^2 =\frac94 \;\Omega^2\:,\hs
M_1^2 =\Omega^2\left( \frac94-12\xi\right)+
\left(1-6\xi\right)6\xi\alpha^2\phi^2,
\hs
M_2^2=\frac14\;\Omega^2, 
\label{2n}\eeq   
and, according to Eq.~(\ref{VT123}), the temperature dependent part of
the effective potential becomes 
\beq
V^{(1)}_T =
\frac{3\ze_R(3)\Om T^3}{2\pi}
\aq1+\frac{8\xi(1-6\xi)\alpha^2\phi^2}{3\Omega^2}\cq^{\frac12}
-\frac{T^2}{2}\xi(1-6\xi)\alpha^2\phi^2+ O(\xi^{5/2})+const 
\:,\label{3n}
\eeq
where the last term is independent on scalar field and can be omitted.

In order to explore the phase transition conditions (\ref{phase}), 
it is convenient to introduce the dimensionless quantities
\beq
\chi^2=\frac{\phi^2}{\mu^2}\:,\hs
\hat T=\frac{T}{\hat\mu}\:,\hs
\hat\Om=\frac{\Om}{\hat\mu}
\:,\eeq
where $\hat\mu^2=\xi(1-6\xi)\alpha^2\mu^2$  and see the
one loop effective potential as a function of $\chi$. 
In the approximation that we are using it reads
\beq
\hat V_{eff}(\chi, \hat T, \hat\Om)&=
&\frac{V_{eff}(\chi)-V_{eff}(0)}{\hat\mu^4}
\nn\\ &\sim& 
\frac{9}{16\pi^2}\;\chi^4 \left[ \ln\chi^2 - \frac{25}{6} \right]
-\frac{3\ze_R(3)\hat\Om\hat T^3}{2\pi}
\aq1-\at1+\frac{8\chi^2}{3\hat\Omega^2}\ct^{\frac12}\cq
-\frac{\hat T^2\chi^2}{2}
\:.\label{4n}
\eeq
Remarcably the arbitrariness related with $\al$ drops from the potential.

\begin{figure}[h]
\begin{center}
\leavevmode
\epsfxsize=7cm
\epsfbox{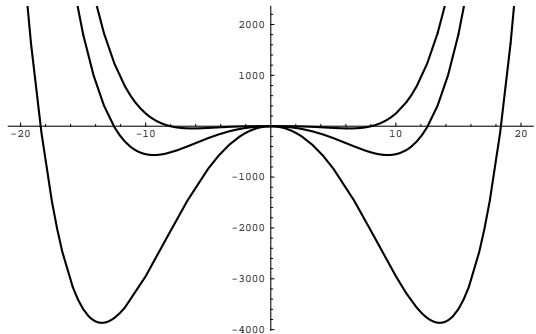}
\end{center}
\vspace{-10ex}
\caption{$\hat V_{eff}(\chi,\hat T,0)$ for 
$\hat T=0$, $\hat T=4$ and $\hat T=8$ 
respectively (from the top)}
\label{Fig1}
\end{figure}

The general Eqs.~(\ref{phase}) are quite complicated and that is why
we start the physical analysis with the particular case of the flat
background metric.
For this case, the  effective potential
\beq
\hat V_{eff}(\chi, \hat T, 0)&=&
\frac{9}{16\pi^2}\;\chi^4 \left[ \ln\chi^2 - \frac{25}{6} \right]
-\frac{\hat T^2\chi^2}{2}
\label{5n}
\eeq
has qualitatively the same form as the temperature dependent effective
potential in an ordinary field theory like scalar QED \cite{dola74-9-3320}.
The temperature contributions lead to the appearance of the positive mass 
term which can provide the phase transition. In our case, however, since the
original theory was massless, the vacuum corresponds to broken symmetry.
The second condition (\ref{phase}) gives the following relations for 
$\chi_c,T_c$ and the value of potential in the points of minima.
\beq
\frac{9}{4\pi^2}\;{\chi_c}^2
\left[ \ln{\chi_c}^2-\frac{11}3\right]= \hat T_c,
\nn\eeq
\beq
\hat V_{eff}(\chi_c, \hat T_c, 0) =
-\frac{1}{4}\;{{\hat T_c}^2}\chi_c^2 
-\frac{9}{32\pi^2} \chi_c^4 < 0
\:,\label{6n}\eeq
\beq
\hat V''_{eff}(\chi_c,\hat T_c, 0) 
=2T_c^2+\frac9{2\pi^2}\chi_c^2 > 0
\nn\:.\eeq
The temperature dependent terms have soft behaviour at
large values of $\phi$ and therefore the $\phi^4$ terms from 
$V_{cl}+V_0^{(1)}$ turn out to be relevant, whereas for the $\phi^2$ terms 
the temperature dependent part is dominating.  The analysis of the potential 
(\ref{6n}) is straightforward.
One can see that in the flat-space, both 
broken and unbroken phases are possible, 
depending on the parameter of normalization $\mu$.
The second order phase transition occurs if $\chi >> 1$ (about two orders)
that corresponds to the big value of the order parameter. 
In this case the magnitude of effective potential -- 
the induced cosmological constant -- does not 
depend on $\mu$ whereas the critical values of $\chi$ and
$T$ do. The effective potential for this case is shown in 
Figure~\ref{Fig1}.
On the other hand, from a physical point of view it is more natural to
consider $\mu$ as a character value of the order parameter $\chi$. Then
one has to put $\chi$ close to $1$ and the phase is unbroken (see 
the first line on Figure~\ref{Fig1}).

\begin{figure}[h]
\begin{center}
\leavevmode
\epsfxsize=7cm
\epsfbox{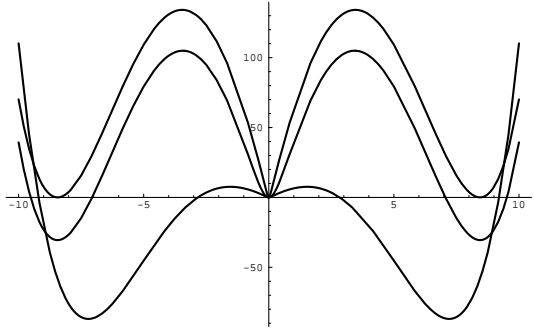}
\end{center}
\vspace{-10ex}
\caption{$\hat V_{eff}(\chi,\hat T,\hat \Om)$ 
for $(\hat T,\hat\Om)=(4.333,0.075)$,
$(\hat T,\hat\Om)=(4.333,0.75)$ and $(\hat T,\hat\Om)=(2.16,0.075)$
respectively (from the top)}
\label{Fig2}
\end{figure}

The conditions (\ref{phase}) for the general nontrivial AdS 
metric are indeed much more complicated and we have analysed them 
numerically. It is convenient to distinguish the two regions
$\hat T<\hat T_0$ and $\hat T>\hat T_0$ ($\hat T_0\simeq4.3$). 
In fact, in the first region $\hat T<\hat T_0$, there is always  
a second order phase transition for any value of $\hat\Om$. 
The depth of the potential hole increase with $\Om$
(remember that, to be consistent with our approximation, 
we have to choose $\hat\Om<<\hat T$).
For $\hat T=\hat T_0$,  a first order phase transition appears for 
$\hat\Om=\hat\Om_0\simeq0$ and of course, a second order
phase transition for any $\Om>\Om_0$.
In the second region, that is $\hat T>\hat T_0$, we can have both first 
or second order phase transitions, depending on the parameters, 
but the condition $\hat\Om<<\hat T$, notably limits the possible 
values of $\hat T$. In fact, one can easily verify that for 
$\hat T=4.5$ the first order phase transition appears for 
$\hat\Om=0.54\sim 0.12 \hat T$. Then, in our approximations, we have 
first order phase transitions for $4.3<\hat T<4.5$ and second order 
phase transitions for $\hat T<4.5$ (see Figure~\ref{Fig2}).

Thus one can meet three different phases depending on the value of $\mu$.
In particular, for $\chi>>1$ and very small values of $\Omega$ there is always 
broken phase but the potential in the point of minima is negative.
Thus we have the second order phase transition and the result is 
qualitatively the same which takes place in the more 
simple $\Om = 0$ case. The situation
for the natural choice of normalization is also similar to the one 
discussed above -- the phase is actually unbroken if we deal with the 
region $T >> \Om$ corresponding to our approximation.  
Now, if we increase the value of $\Om$ the qualitative result becomes 
different but still depends on the value of the normalization constant $\mu$. 
As in the previous case, the phase can be broken only if 
$\mu$ is about two orders less than $\chi$.
The effect of negative curvature, even if it is very small, is relevant in the
vicinity of $\phi = 0$ because it changes the sign of the second derivative
of the potential and therefore leads to the stability of this point.

\section{Conclusion}

We have considered the quantum theory of conformal factor coupled to
the Higgs field in the high temperature regim. It turns out that
the temperature effects lead to drastical change in the effective potential
of the Higgs field, as a result the quantum corrections dominate in
the effective potential that was not the case in the zero-temperature 
approximation \cite{foo}.
The temperature dependent terms have soft behaviour at
large values of $\phi$ and therefore the $\phi^4$ terms from 
$V_{cl}+V_0^{(1)}$ turn out to be relevant, whereas for the $\phi^2$ terms 
the temperature dependent part is dominating.  In this respect the picture is
qualitatively the same as for ordinary scalar field theory in the 
high-temperature regime \cite{dola74-9-3320}.
However there are crucial differences between the two theories. In the theory 
under consideration the quantum effects are dominating because of
the specific form of the renormalization group flows (\ref{17}) which end
in the IR free point. Consequently the scalar field gets rescaled and the
rescaled quantity plays a role of the order parameter.
Simultaneously the classical coupling constant $f$ 
vanishes in the fixed point much faster
than the nonminimal parameter and the "new" scalar self-coupling is 
expressed in terms of the nonminimal parameter and scaling constant.

The effective potential is given by a 
very complicated expression and we have analysed it in the vicinity
of the IR stable fixed point in the regim of high temperature and
small negative curvature. 
The second order phase transition takes place for most of the values of the 
parameters of the theory. For some special values it takes place the first 
order phase transition too.

\ack{Authors thank I.L. Buchbinder, I.V. Tyutin and L. Vanzo for 
useful discussions. 
I.Sh. is grateful to Dipartimento di Fisica, Universit\`a di Trento
for kind hospitality and INFN for the support for visiting Trento,
where this work has been started.
The work of I.Sh. was also supported in part by MEC-DGICYT, Spain.}

\end{document}